\pdfoutput=1
\documentclass[12pt,a4paper]{article}
\usepackage{ifthen} 

\newboolean{pdflatex}
\setboolean{pdflatex}{true} 

\newboolean{articletitles}
\setboolean{articletitles}{true} 

\newboolean{uprightparticles}
\setboolean{uprightparticles}{false} 

\newboolean{inbibliography}
\setboolean{inbibliography}{false} 

\usepackage[top=1in, bottom=1.25in, left=1in, right=1in]{geometry}

\usepackage{microtype}
\usepackage{lineno}  
\usepackage{xspace} 
\usepackage{caption} 

\usepackage{graphicx}  
\usepackage{color}
\usepackage{colortbl}
\graphicspath{{./figs/}} 

\usepackage{amsmath} 
\usepackage{amssymb}
\usepackage{amsfonts}
\usepackage{upgreek} 

\newcommand*\patchAmsMathEnvironmentForLineno[1]{%
\expandafter\let\csname old#1\expandafter\endcsname\csname #1\endcsname
\expandafter\let\csname oldend#1\expandafter\endcsname\csname
end#1\endcsname
 \renewenvironment{#1}%
   {\linenomath\csname old#1\endcsname}%
   {\csname oldend#1\endcsname\endlinenomath}%
}
\newcommand*\patchBothAmsMathEnvironmentsForLineno[1]{%
  \patchAmsMathEnvironmentForLineno{#1}%
  \patchAmsMathEnvironmentForLineno{#1*}%
}
\AtBeginDocument{%
\patchBothAmsMathEnvironmentsForLineno{equation}%
\patchBothAmsMathEnvironmentsForLineno{align}%
\patchBothAmsMathEnvironmentsForLineno{flalign}%
\patchBothAmsMathEnvironmentsForLineno{alignat}%
\patchBothAmsMathEnvironmentsForLineno{gather}%
\patchBothAmsMathEnvironmentsForLineno{multline}%
\patchBothAmsMathEnvironmentsForLineno{eqnarray}%
}

\usepackage{hyperref}    
\usepackage[all]{hypcap} 

\usepackage{xspace}
\usepackage{upgreek}

\def\lhcb {\mbox{LHCb}\xspace}

\def\babar  {\mbox{BaBar}\xspace}
\def\belle  {\mbox{Belle}\xspace}

\def\MagUp {\mbox{\em Mag\kern -0.05em Up}\xspace}

\ifthenelse{\boolean{uprightparticles}}%
{

 \def\Ppi         {\ensuremath{\uppi}\xspace}

 \def\PDelta      {\ensuremath{\Delta}\xspace}
 \def\PXi      {\ensuremath{\Xi}\xspace}
 \def\PLambda      {\ensuremath{\Lambda}\xspace}
 \def\PSigma      {\ensuremath{\Sigma}\xspace}
 \def\POmega      {\ensuremath{\Omega}\xspace}
 \def\PUpsilon      {\ensuremath{\Upsilon}\xspace}

 \def\PB      {\ensuremath{\mathrm{B}}\xspace}
 
 \def\PD      {\ensuremath{\mathrm{D}}\xspace}

 \def\PK      {\ensuremath{\mathrm{K}}\xspace}

 \def\Pb      {\ensuremath{\mathrm{b}}\xspace}

 \def\Pi      {\ensuremath{\mathrm{i}}\xspace}

 \def\Ps      {\ensuremath{\mathrm{s}}\xspace}

}
{

 \def\Ppi         {\ensuremath{\pi}\xspace}

 \mathchardef\PDelta="7101
 \mathchardef\PXi="7104
 \mathchardef\PLambda="7103
 \mathchardef\PSigma="7106
 \mathchardef\POmega="710A
 \mathchardef\PUpsilon="7107
 
 \def\PB      {\ensuremath{B}\xspace}
 
 \def\PD      {\ensuremath{D}\xspace}

 \def\PK      {\ensuremath{K}\xspace}

 \def\Pb      {\ensuremath{b}\xspace}

 \def\Pi      {\ensuremath{i}\xspace}

 \def\Ps      {\ensuremath{s}\xspace}

}

\makeatletter
\ifcase \@ptsize \relax
  \newcommand{\miniscule}{\@setfontsize\miniscule{4}{5}}
\or
  \newcommand{\miniscule}{\@setfontsize\miniscule{5}{6}}
\or
  \newcommand{\miniscule}{\@setfontsize\miniscule{5}{6}}
\fi
\makeatother

\DeclareRobustCommand{\optbar}[1]{\shortstack{{\miniscule (\rule[.5ex]{1.25em}{.18mm})}
  \\ [-.7ex] $#1$}}

\def\squark    {{\ensuremath{\Ps}}\xspace}

\def\bquark    {{\ensuremath{\Pb}}\xspace}

\def\pion   {{\ensuremath{\Ppi}}\xspace}
\def\piz    {{\ensuremath{\pion^0}}\xspace}

\def\pip    {{\ensuremath{\pion^+}}\xspace}
\def\pim    {{\ensuremath{\pion^-}}\xspace}

\def\kaon    {{\ensuremath{\PK}}\xspace}
  \def\Kbar    {{\kern 0.2em\overline{\kern -0.2em \PK}{}}\xspace}

\def\KorKbar    {\kern 0.18em\optbar{\kern -0.18em K}{}\xspace}

\def\Kp      {{\ensuremath{\kaon^+}}\xspace}
\def\Km      {{\ensuremath{\kaon^-}}\xspace}

\def\KS      {{\ensuremath{\kaon^0_{\mathrm{ \scriptscriptstyle S}}}}\xspace}

  \def\Dbar    {{\kern 0.2em\overline{\kern -0.2em \PD}{}}\xspace}
\def\D       {{\ensuremath{\PD}}\xspace}
\def\Db      {{\ensuremath{\Dbar}}\xspace}
\def\DorDbar    {\kern 0.18em\optbar{\kern -0.18em D}{}\xspace}
\def\Dz      {{\ensuremath{\D^0}}\xspace}
\def\Dzb     {{\ensuremath{\Dbar{}^0}}\xspace}

\def\Dstarzb {{\ensuremath{\Dbar{}^{*0}}}\xspace}
\def\Dstarzbpar {{\ensuremath{\Dbar{}^{(*)0}}}\xspace}

\def\B       {{\ensuremath{\PB}}\xspace}
\def\Bbar    {{\ensuremath{\kern 0.18em\overline{\kern -0.18em \PB}{}}}\xspace}

\def\BorBbar    {\kern 0.18em\optbar{\kern -0.18em B}{}\xspace}
\def\Bz      {{\ensuremath{\B^0}}\xspace}

\def\Bzb     {{\ensuremath{\Bbar{}^0}}\xspace}

\def\Bpm     {{\ensuremath{\B^\pm}}\xspace}

\def\Bd      {{\ensuremath{\B^0}}\xspace}
\def\Bs      {{\ensuremath{\B^0_\squark}}\xspace}

  \def\Y#1S{\ensuremath{\PUpsilon{(#1S)}}\xspace}

\def\Lbar        {{\ensuremath{\kern 0.1em\overline{\kern -0.1em\PLambda}}}\xspace}
\def\LorLbar    {\kern 0.18em\optbar{\kern -0.18em \PLambda}{}\xspace}

\def\BF         {{\ensuremath{\mathcal{B}}}\xspace}

\def\BR         {\BF}
\newcommand{\decay}[2]{\ensuremath{#1\!\to #2}\xspace}         

\def\to                 {\ensuremath{\rightarrow}\xspace}

\def\order   {{\ensuremath{\mathcal{O}}}\xspace}

\def\CP                {{\ensuremath{C\!P}}\xspace}

\def\AT#1     {\ensuremath{A_{\mathrm{T}}^{#1}}\xspace}           

\def\C#1      {\ensuremath{\mathcal{C}_{#1}}\xspace}                       
\def\Cp#1     {\ensuremath{\mathcal{C}_{#1}^{'}}\xspace}                    
\def\Ceff#1   {\ensuremath{\mathcal{C}_{#1}^{\mathrm{(eff)}}}\xspace}        
\def\Cpeff#1  {\ensuremath{\mathcal{C}_{#1}^{'\mathrm{(eff)}}}\xspace}       
\def\Ope#1    {\ensuremath{\mathcal{O}_{#1}}\xspace}                       
\def\Opep#1   {\ensuremath{\mathcal{O}_{#1}^{'}}\xspace}                    

\newcommand{\tev}{\ifthenelse{\boolean{inbibliography}}{\ensuremath{~T\kern -0.05em eV}}{\ensuremath{\mathrm{\,Te\kern -0.1em V}}}\xspace}
\newcommand{\gev}{\ensuremath{\mathrm{\,Ge\kern -0.1em V}}\xspace}
\newcommand{\mev}{\ensuremath{\mathrm{\,Me\kern -0.1em V}}\xspace}
\newcommand{\kev}{\ensuremath{\mathrm{\,ke\kern -0.1em V}}\xspace}
\newcommand{\ev}{\ensuremath{\mathrm{\,e\kern -0.1em V}}\xspace}
\newcommand{\gevc}{\ensuremath{{\mathrm{\,Ge\kern -0.1em V\!/}c}}\xspace}
\newcommand{\mevc}{\ensuremath{{\mathrm{\,Me\kern -0.1em V\!/}c}}\xspace}
\newcommand{\gevcc}{\ensuremath{{\mathrm{\,Ge\kern -0.1em V\!/}c^2}}\xspace}
\newcommand{\gevgevcccc}{\ensuremath{{\mathrm{\,Ge\kern -0.1em V^2\!/}c^4}}\xspace}
\newcommand{\mevcc}{\ensuremath{{\mathrm{\,Me\kern -0.1em V\!/}c^2}}\xspace}

\def\invfb   {\ensuremath{\mbox{\,fb}^{-1}}\xspace}

\def\order{{\ensuremath{\mathcal{O}}}\xspace}

\def\gsim{{~\raise.15em\hbox{$>$}\kern-.85em
          \lower.35em\hbox{$\sim$}~}\xspace}
\def\lsim{{~\raise.15em\hbox{$<$}\kern-.85em
          \lower.35em\hbox{$\sim$}~}\xspace}

\def\pt         {\mbox{$p_{\mathrm{ T}}$}\xspace}

\def\tell1  {TELL1\xspace}
\def\ukl1   {UKL1\xspace}

\usepackage{cite} 
\usepackage{mciteplus}

\usepackage{longtable} 

\begin{document}

\renewcommand{\thefootnote}{\fnsymbol{footnote}}
\setcounter{footnote}{1}


\begin{titlepage}
\pagenumbering{roman}

\vspace*{-1.5cm}
\centerline{\large EUROPEAN ORGANIZATION FOR NUCLEAR RESEARCH (CERN)}
\vspace*{1.5cm}
\noindent
\begin{tabular*}{\linewidth}{lc@{\extracolsep{\fill}}r@{\extracolsep{0pt}}}
\ifthenelse{\boolean{pdflatex}}
 & & CERN-EP-2018-158 \\  
 & & LHCb-PAPER-2018-015 \\  
 & & November 8, 2018 \\ 
 & & \\
\end{tabular*}

\vspace*{2.5cm}

{\normalfont\bfseries\boldmath\LARGE
\begin{center}
      Observation of $\Bs \to \Dstarzb \phi$ and search for $\Bd \to \Dzb \phi$ decays
\end{center}
}

\vspace*{1.0cm}

\begin{center}
The LHCb collaboration\footnote{Authors are listed at the end of this paper.}
\end{center}

\vspace{\fill}

\begin{abstract}
  \noindent
The first observation of the $\Bs \to \Dstarzb \phi$ decay is reported, with a significance of more than seven standard deviations, from an analysis of $pp$  collision data corresponding to an integrated luminosity of  3\invfb, collected with the \lhcb detector at centre-of-mass energies of $7$ and $8$~\tev. The branching fraction is measured relative to that of the topologically similar decay $\Bz\to\Dzb \pip\pim$ and is found to be $\BR(\Bs\to\Dstarzb\phi) = (3.7 \pm 0.5 \pm 0.3 \pm 0.2) \times 10^{-5}$, where the first uncertainty is statistical, the second systematic, and the third from the branching fraction of the $\Bz\to\Dzb\pip \pim$ decay. The fraction of longitudinal polarisation in this decay is measured to be ${f_{\rm L} =(73 \pm 15 \pm 4)\%}$. The most precise determination of the branching fraction for the $\Bs\to\Dzb\phi$ decay is also obtained, \mbox{$\BR(\Bs\to\Dzb\phi) = (3.0 \pm 0.3 \pm 0.2 \pm 0.2) \times 10^{-5}$}. An upper limit, \mbox{$\BR(\Bz\to\Dzb\phi) < 2.0 \ (2.3) \times 10^{-6}$ at $90\%$} (95\%) confidence level is set. A constraint on the $\omega-\phi$ mixing angle $\delta$ is set at $|\delta| < 5.2^\circ~ (5.5^\circ)$ at $90\%$ (95\%) confidence level.
\end{abstract}

\vspace*{1.0cm}

\begin{center}
  Published in Phys.~Rev.~D98 (2018) 071103(R)
\end{center}

\vspace{\fill}

{\footnotesize
\centerline{\copyright~2018 CERN for the benefit of the \lhcb collaboration. \href{http://creativecommons.org/licenses/by/4.0/}{CC-BY-4.0} licence.}}
\vspace*{2mm}

\end{titlepage}


\newpage
\setcounter{page}{2}
\mbox{~}

\cleardoublepage


\newcommand{\al}{\ensuremath{\kern 0.5em }}
\newcommand{\all}{\ensuremath{\kern 0.25em }}

\newcommand{\alm}{\ensuremath{\kern -0.5em }}
\newcommand{\allm}{\ensuremath{\kern -0.25em }}

\renewcommand{\thefootnote}{\arabic{footnote}}
\setcounter{footnote}{0}


\pagestyle{plain} 
\setcounter{page}{1}
\pagenumbering{arabic}


The precise measurement of the angle $\gamma$ of the Cabibbo-Kobayashi-Maskawa (CKM) Unitarity Triangle~\cite{PhysRevLett.10.531,PTP.49.652}  is a central topic in flavour physics experiments. Its determination at the subdegree level in tree-level open-charm $b$-hadron decays is theoretically clean~\cite{Brod:2013sga,Brod:2014bfa} and provides a standard candle for measurements sensitive to new physics effects~\cite{CKMfitter2013}. In addition to the results from the  $B$ factories~\cite{Bevan:2014iga}, various measurements from \lhcb~\cite{HFLAV16,LHCb-PAPER-2016-032,LHCb-CONF-2018-002} allow  the angle $\gamma$ to be determined with an uncertainty of around $5^\circ$. However, no single measurement dominates the world average, as the most accurate measurements have an accuracy of $\order(10^\circ-20^\circ)$~\cite{LHCb-PAPER-2017-047,LHCb-PAPER-2018-017}.  Alternative methods are therefore important  to improve the precision.  Among them, an analysis of the decays $\Bs \to \Dstarzbpar\phi$  open possibilities to offer competitive experimental precision on the angle $\gamma$~\cite{Gronau:1990ra,Gronau:2004gt,Gronau:2007bh,Ricciardi:1243068}, where the $\Dstarzb$ meson can be partially reconstructed~\cite{LHCb-PAPER-2017-021}.

\begin{figure}[b]
\centering
\includegraphics[scale=0.55]{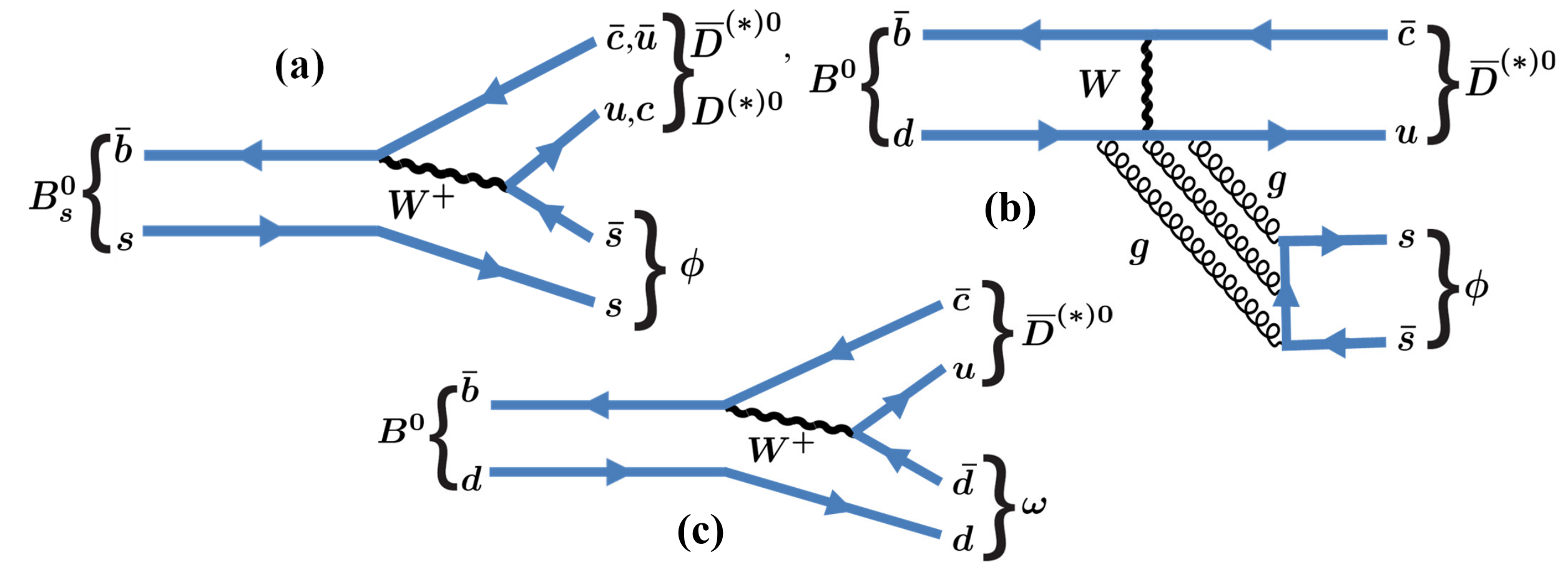}
\caption{Diagrams that contribute to the (a) colour-suppressed $\Bs  \to \Db^{(*)0}/D^{(*)0} \phi$, (b) $W$-exchange OZI-suppressed $\Bd  \to \Dzb/\Dz \phi$ and the (c) colour-suppressed $\Bd  \to \Dzb \omega$ decays.}
\label{fig:Feynman}
\end{figure}

The tree-level Feynman diagrams for the $\Bs \to \Dstarzbpar \phi$ decays are shown in Fig.~\ref{fig:Feynman}~(a). The inclusion of charge-conjugated processes is implied throughout the paper. The decay  $\Bs \to \Dzb \phi$ was first observed  by the \lhcb collaboration~\cite{LHCb-PAPER-2013-035} using a data sample corresponding to an integrated luminosity of  1\invfb, while no prior results exist for $\Bs \to \Dstarzb \phi$ decays. The branching fraction $\BR(\Bs \to \Dzb \phi)$ is $(3.0 \pm 0.8) \times 10^{-5}$~\cite{LHCb-PAPER-2013-035, PDG2018}. The $\Bs \to \Dstarzb \phi$ decay is a vector-vector mode and can proceed through different polarisation amplitudes. A measurement of its fraction of longitudinal polarisation ($f_{\rm L}$) is of particular interest because a significant deviation from unity would confirm previous results from similar colour-suppressed $\Bd$ decays~\cite{Lees:2011gw,Matvienko:2015gqa}, as expected from theory~\cite{Blechman:2004vc,Beneke:2005we}. This also helps to constrain QCD models and to search for effects of physics beyond the Standard Model (see review on polarisation in $B$ decays in Ref.~\cite{PDG2018}).

The $\Bd\to\Dzb\phi$ decay can proceed by leading-order Feynman diagrams shown either in Fig.~\ref{fig:Feynman}~(b) or in Fig.~\ref{fig:Feynman}~(c), followed by $\omega-\phi$ mixing.
The $W$-exchange decay is suppressed by the Okubo-Zweig-Iizuka (OZI) rule~\cite{Okubo_1963fa,Zweig_1964jf,Iizuka_1966fk}. Assuming that the colour-suppressed $\Bd  \to \Dzb \omega$ decay dominates,
the  branching fraction of $\Bd\to\Dzb\phi$ is predicted and can be used to determine the mixing angle $\delta$~\cite{Gronau:2008kk}. The relation between the branching fractions and mixing angle can be written as \mbox{$\tan^2\delta = {\BR(\Bz\to\Dzb\phi)}/{\BR(\Bz\to\Dzb\omega)} \times {\Phi(\omega)}/{\Phi(\phi)}$}, where $\Phi(\omega)$ and $\Phi(\phi)$ are the integrals of the phase-space factors computed over the resonant lineshapes. A calculation, using a recent result on  $\BR(\Bd\to\Dzb \omega)$~\cite{Lees:2011gw} and taking into account phase-space factors, gives $\BR(\Bd \to\Dzb \phi) = (1.6\pm0.1)\times 10^{-6}$. The ratio ${\Phi(\omega)}/{\Phi(\phi)} =1.05 \pm 0.01$ is used, where the uncertainty comes from the limited knowledge on the shape parameters of the two resonances. The previous experimental upper limit on this branching fraction was $\BR(\Bd \to\Dzb \phi)<11.7\times 10^{-6}$ at $90\%$ confidence level (CL)~\cite{Aubert:2007nw}. The new measurement presented in this Letter also allows the $\omega-\phi$ mixing angle to be determined~\cite{Gronau:2008kk,Benayoun:2007cu}.

In this Letter, results on  the $B^0_{(s)} \to  \Dstarzbpar\phi$ decays are presented, where the $\phi$ meson is reconstructed through its decay to a $\Kp \Km$ pair and the $\Dzb$ meson decays to $\Kp \pim$. The $\Bs \to \Dstarzb \phi$ decay is partially reconstructed without inclusion of the neutral pion or photon from the $\Dstarzb$ meson decay. The analysis is based on a data sample corresponding to $3.0  \,{\rm fb}^{-1}$ of integrated luminosity, of which approximately one third (two thirds) were collected by the LHCb detector from $pp$ collisions at a centre-of-mass energy of $7 \ (8) \ \tev$.

The \lhcb detector is a single-arm forward spectrometer covering the \mbox{pseudorapidity} range $2<\eta <5$, described in detail in Refs.~\cite{Alves:2008zz,LHCb-DP-2014-002}. The online event selection is performed by a trigger~\cite{LHCb-DP-2012-004}, which consists of a hardware stage, based on information from the calorimeter and muon systems, followed by a software stage, which applies a full event reconstruction and requires a two-, three- or four-track secondary vertex with a large sum of the component of the momentum transverse to the beam, \pt, of the tracks and a significant displacement from all primary $pp$-interaction vertices~(PV).

The selection requirements for the $B^0_{(s)} \to  \Dstarzbpar\phi$ signals are the same as those used for the branching fraction measurements of $B_{(s)}^0 \to \Dzb \Kp \Km$, as described in detail in Ref.~\cite{LHCb-PAPER-2018-014}. The selection criteria are optimised using the \mbox{$\Bd\to\Dzb\pip \pim$} decay as a normalisation channel. Signal  $B^0_{(s)} \to \Dzb \Kp \Km$ candidates are formed by combining  \Dzb candidates, reconstructed in the final states $\Kp\pim$, with two additional particles of opposite charge, identified as kaons, whose tracks are required to be inconsistent with originating from a PV. They must have sufficiently high $p$ and $\pt$ and be within the fiducial acceptance of the two ring-imaging Cherenkov detectors~\cite{LHCb-DP-2012-003} used for particle identification (PID) of charged hadrons. The \Dzb decay products are required to form a good quality vertex with an invariant mass within 25~\mevcc of the known \Dzb mass~\cite{PDG2018}. The \Dzb and two kaon candidates must form a good vertex. The reconstructed \Dzb and $B$ vertices are required to be significantly displaced from any PV. To improve the $B$-candidate invariant-mass resolution, a kinematic fit~\cite{Hulsbergen:2005pu} is used, constraining the \Dzb candidate invariant mass to its known value~\cite{PDG2018} and the $B$ momentum to point back to the PV with smallest $\chi^2_{\textrm{IP}}$, where $\chi^2_{\textrm{IP}}$ is defined as the difference in the vertex-fit $\chi^2$ of a given PV reconstructed with and without the particle under consideration. By requiring the reconstructed \Dzb vertex to be displaced downstream from the reconstructed \Bz vertex, backgrounds from both charmless $B$ decays and charmed mesons produced at the PV are reduced to a negligible level.
Background from $\Bz \to D^{*}(2010)^{-} \Kp$ decays is removed by requiring the reconstructed mass difference $m_{\Dzb\pim}-m_{\Dzb}$ not to be within $\pm4.8$~\mevcc of its known value~\cite{PDG2018} after assigning the pion mass to the kaon. To further distinguish signal from combinatorial background, a multivariate analysis based on a Fisher discriminant~\cite{Fisher:1936et} is applied. The discriminant is optimised by maximising the statistical significance of  $\Bd\to\Dzb\pip \pim$ candidates selected in a similar way. The discriminant uses the following information: the smallest values of $\chi^2_{\rm IP}$ and $\pt$ of the prompt tracks from the $B$-decay vertex; the $B$ flight-distance significance; the $D~\chi^2_{\rm IP}$, and the signed minimum cosine of the angle between the direction  of one of the prompt tracks from the $B$ decay and the $\Dzb$ meson, as projected in the plane perpendicular to the beam axis.

Candidate $B^0_{(s)} \to \Dzb \Kp \Km$ decays with invariant masses in the range \mbox{[5000, 6000]}~\mevcc are retained. After all selection requirements are applied, less than $1\%$ of the events contain multiple candidates, and a single candidate is chosen based on the fit quality of the $B$- and $D$-meson vertices and on the PID information of the $\Dzb$ decay products.
The effect due to the multiple candidate selection is negligible~\cite{Koppenburg:2017zsh}.

\begin{figure}[t]
\begin{center}
\includegraphics[scale=0.53]{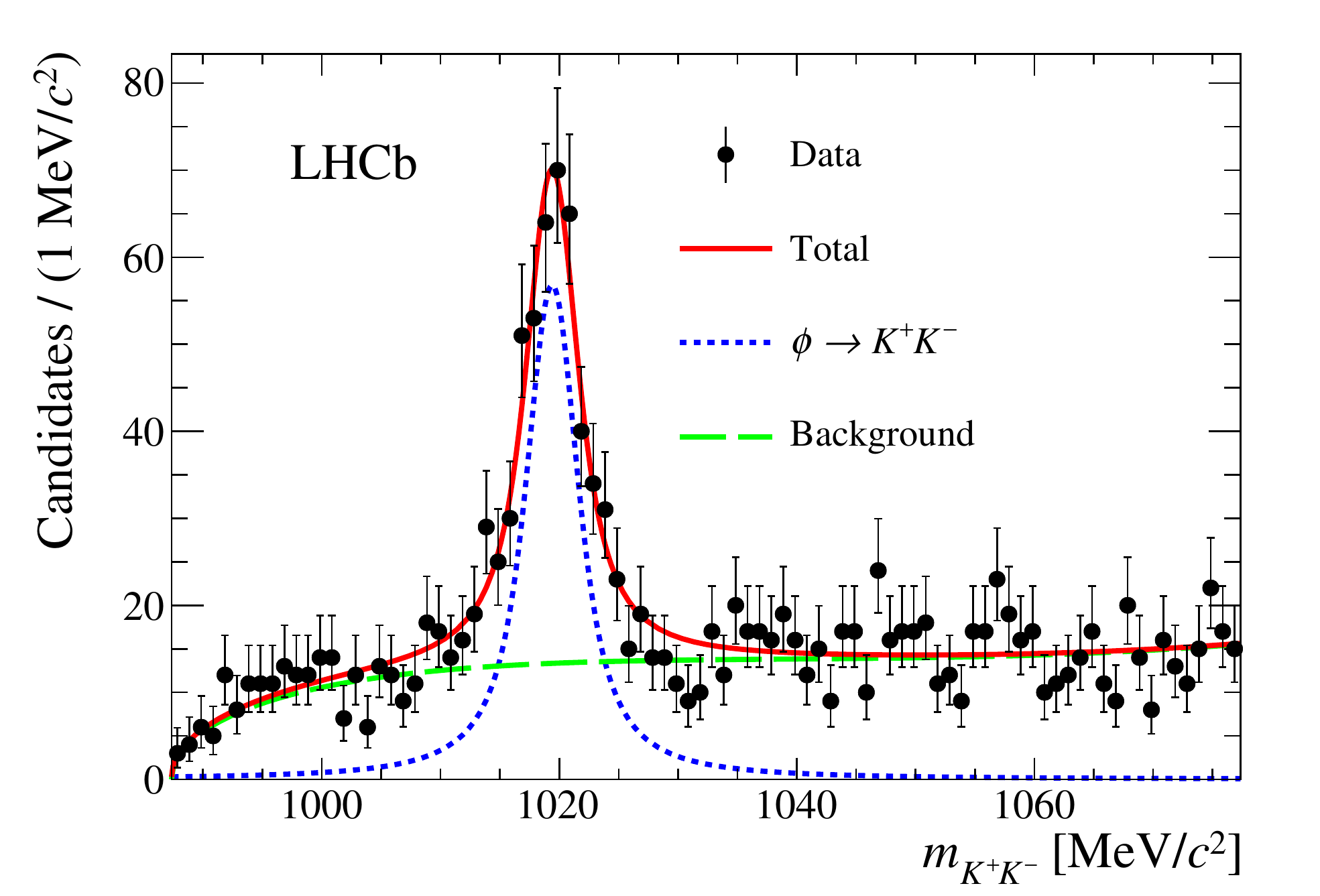}
\caption{Fit to the $m_{\Kp \Km}$ invariant-mass distribution. Data points are shown in black, the fitted total  PDF as a solid (red) line and the component PDFs as dashed lines: (green) background and (blue) signal.}
\label{Plot_BstoDPhi_m23_Per8MeV}
\end{center}
\end{figure}

The distribution of the invariant mass of the $\Kp\Km$ pair, $m_{\Kp\Km}$, shown in Fig.~\ref{Plot_BstoDPhi_m23_Per8MeV}, is obtained from a narrow window, ${[2m_K, \ 2m_K + 90 \ \mevcc]}$, covering the $\phi$ meson mass~\cite{PDG2018} and where $m_K$ is the known kaon mass. An extended unbinned maximum-likelihood fit to the invariant-mass distribution of the $\phi$ candidates, $m_{\Kp \Km}$, is performed to statistically separate $\phi$ signal from background by means of the \textit{sPlot} technique~\cite{Xie:2009rka, Pivk:2004ty}. The $\phi$ meson invariant-mass distribution is modelled with a Breit--Wigner probability density function (PDF) convolved with a Gaussian resolution function. The width of the Breit-Wigner function is fixed to the known $\phi$ width~\cite{PDG2018}.
The PDF for the background is a phase space factor $p \times q$ multiplied by a quadratic function $(1+ax+b(2x^2-1))$, where $p$ and $q$ are the momentum of the kaon in the $\Kp\Km$ rest frame and the momentum of the $\Dzb$ in the $\Dzb\Kp\Km$ rest frame, respectively. The variable $x$ is defined as $2\times (m_{\Kp \Km}-2m_K)/\Delta -1$, where $\Delta$ is the width of the $m_{\Kp \Km}$ mass window so that $x$ is in the range $[-1, \ 1]$. The parameters $a$ and $b$ are free to vary in the fit. The fit describes the data well ($\chi^2/{\rm ndf}=61/82$).
The yields determined by the fit are $427\pm 30$ for the $\phi \to \Kp\Km$ decay and $1152\pm 41$ for the background.

\begin{figure}[t]
\begin{center}
\includegraphics[scale=0.53]{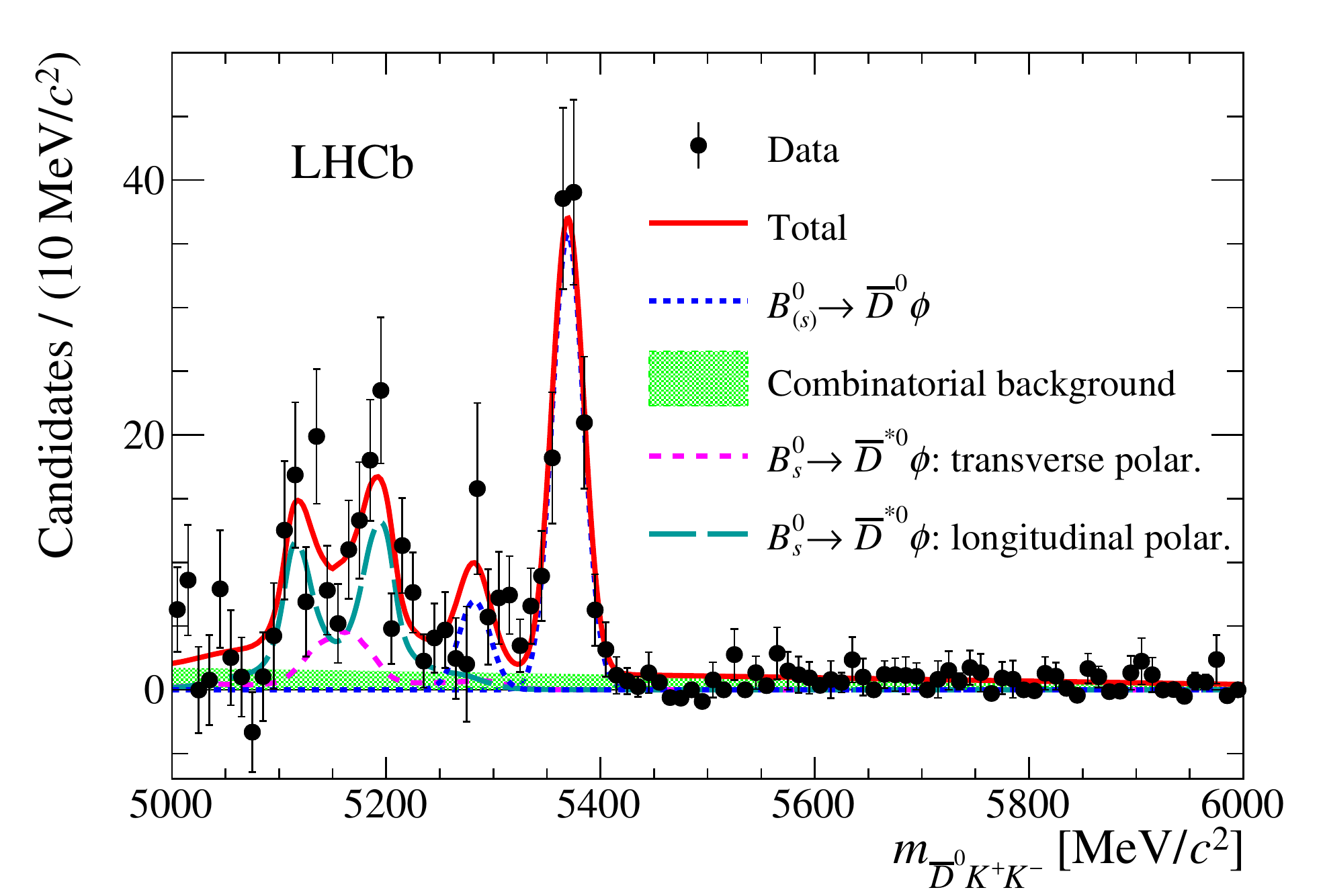}
\caption{Fit to the $m_{\Dzb \Kp \Km}$ invariant-mass distribution of $\Dzb \phi$ candidates obtained using the \textit{sPlot} technique. Data are shown as black points. The total fit function is displayed as a red solid line and the different contributions are represented as dashed lines and shadowed area: (blue short dashed) the $\Bs\to\Dzb\phi$ and $\Bz\to\Dzb\phi$ signal decays,  the $\Bs\to\Dstarzb\phi$ signal decay, with (cyan long dashed) longitudinal and (pink middle dashed) transverse polarisation and (green shaded area) the combinatorial background. }
\label{Plot_BstoDPhi_BM_Phi_sw_with_B02D0Phi_sWprocess_Per8MeV}
\end{center}
\end{figure}

Figure~\ref{Plot_BstoDPhi_BM_Phi_sw_with_B02D0Phi_sWprocess_Per8MeV} displays the $sPlot$-projected invariant-mass distribution of $\Dzb \Kp\Km$, $m_{\Dzb \Kp\Km}$, of $B^0_{(s)} \to  \Dstarzbpar\phi$ candidates. The $m_{\Kp \Km}$ invariant mass is used as the discriminating variable and it is only weakly correlated with the $m_{\Dzb\Kp \Km}$ invariant mass (less than $6\%$). A $\Bs \to \Dzb\phi$ signal peak is visible at the $\Bs$ mass, while there is a statistically insignificant excess of $\Bd\to\Dzb\phi$ candidates at the $\Bd$ mass. In the region below $m_\Bs-m_\piz$ (up to resolution effects), a wider structure is visible and can be attributed to the vector-vector decay $\Bs \to \Dstarzb [\to\Dzb\piz/\Dzb\gamma]\phi$.

An extended unbinned maximum-likelihood fit is performed to determine the number of $\Bd$ and $\Bs$ decaying into the $\Dzb \phi$ final state and that of the mode $\Bs \to \Dstarzb\phi$ together with the value of the longitudinal polarisation fraction $f_{\rm L}$. The $\Bs \to \Dzb \phi$ mode is modelled by a Gaussian function, for which the mean value and resolution are free parameters.
The $\Bd$ signal is modelled by a Gaussian function with the same resolution as the $\Bs$ mode and a mean constrained with respect to that of the $\Bs$ signal using the known $m_\Bs-m_\Bd$ mass difference~\cite{PDG2018}. The $\Bs \to \Dstarzb \phi$ signal is modelled by non-parametric PDFs, built from large simulated samples, using a kernel estimation technique~\cite{Cranmer:2000du}.
Its shape, as a function of the $\Dzb \Kp \Km$ invariant-mass distribution, strongly depends on the polarisation of the decay amplitude. Two extreme polarisation configurations are considered: fully longitudinal ($f_{\rm L}=1$) or transverse ($f_{\rm L}=0$). A global PDF for each polarisation (${\cal P}_{\rm long/trans}$) is obtained as the average of the PDF of the two decays $\Dstarzb\to\Dzb \piz/\Dzb\gamma$, weighted according to their relative branching fraction~\cite{PDG2018}. The total PDF for the $\Dstarzb\phi$ signal is then modelled as the sum $f_{\rm L} \times {\cal P}_{\rm long} + (1- f_{\rm L})\times {\cal P}_{\rm trans}$. The residual background is accounted for with a first-order polynomial function. The yields obtained from this fit are $N_{\Bs\to\Dzb\phi} = 132 \pm 13$, $N_{\Bd\to\Dzb\phi} = 26 \pm 11$, and $N_{\Bs\to\Dstarzb\phi} = 163 \pm 19$, with $f_{\rm L} = (73 \pm 15)\%$.

The branching fractions of $\B^0_{(s)} \to \Dstarzbpar \phi$ are measured as
\begin{equation}
    	\frac{\BR(\B^0_{(s)} \to \Dstarzbpar \phi)}{\BR(\Bz\to\Dzb \pip \pim)} = \frac{N_{\B^0_{(s)} \to \Dstarzbpar  \phi} \times \varepsilon(\Bz\to \Dzb \pip \pim)}{N_{\Bz\to\Dzb \pip \pim} \times \varepsilon(\B^0_{(s)} \to \Dstarzbpar  \phi)} \times \frac{\mathcal F}{\BR(\phi\to \Kp \Km)},
    	\label{equ:Rapport_BR_BDPhi_OVER_B02DPiPi}
\end{equation}
where $\mathcal F$ is 1 for $\Bd$ decays and $f_d/f_s$ for $\Bs$ decays. In this ratio, the ratio between the signal and normalisation modes is required.
The efficiency and the number of selected signals for the normalisation mode are: $\varepsilon(\Bd\to\Dzb\pip \pim) = (10.6 \pm 0.3)\times 10^{-4}$ and $N_{\Bd\to\Dzb\pip \pim} = 29 \ 940 \pm 240$ (see Ref.~\cite{LHCb-PAPER-2018-014} for details). The efficiency includes various effects related to reconstruction, triggering and selection of the signal events.
Efficiencies are determined from simulation with data-driven corrections applied. The efficiencies of the modes $\B^0_{s} \to \Dzb \phi$ and $\B^0 \to \Dzb \phi$ are statistically consistent and are equal to \mbox{$\varepsilon(B^0_{(s)}\to\Dzb\phi) = (11.1 \pm 0.3)\times 10^{-4}$}. For the $\Bs \to \Dstarzb \phi$ decay, the efficiency is obtained as the average of the four following sets of simulated events: fully transverse/longitudinal decays with the decays $\Dstarzb\to\Dzb \piz/\Dzb\gamma$, where the obtained \mbox{$f_{\rm L}=(73\pm15)\%$} and the branching fractions of the $\Dstarzb$ sub-decays are used. The efficiency, after data corrections, is found to be $\varepsilon(B_{s}\to\Dstarzb \phi) = (10.8 \pm 0.1)\times 10^{-4}$.

In the fit to the $m_{\Kp\Km}$ distribution, the background is modelled by a single set of parameters $a$ and $b$. However, the background receives contributions from broad $\Kp\Km$ $S$-wave amplitudes, which could be different for the various $\B^0_{(s)} \to \Dstarzbpar \Kp\Km$ modes. Since a full amplitude analysis is beyond the scope of this measurement, the following study is performed: the candidates shown in Fig.~\ref{Plot_BstoDPhi_m23_Per8MeV} are divided into three subsamples: $\Bs \to \Dstarzbpar\phi$-like candidates with $m_{\Dzb \Kp\Km} \in [5000, 5240] \cup [5310, 5400]$~\mevcc, \mbox{$\Bd \to \Dzb \phi$-like} candidates with $m_{\Dzb \Kp\Km} \in [5240, 5310]$~\mevcc, and combinatorial background candidates with $m_{\Dzb \Kp\Km}$ above $5400$~\mevcc.
The parameters $a$ and $b$  of the quadratic background function are determined independently for the three subsamples and are found to be consistent with each other. Using the results from the fits to the three subsamples to describe the $\Kp\Km$ background, pseudoexperiments are generated to produce $\Dzb\Kp\Km$ samples that mimic the data. The signal PDF for the $B^0_{(s)} \to  \Dstarzbpar\phi$ decays and the PDFs for various \bquark-hadron decays are taken from the nominal fit to $m_{\Dzb \Kp \Km}$ as described in  Ref.~\cite{LHCb-PAPER-2018-014} are considered. The fits to the $m_{\Kp\Km}$ and $m_{\Dzb\phi}$ distributions are then repeated to determine the pull distributions of $N_{\Bs\to\Dzb\phi}$, $N_{\Bd\to\Dzb\phi}$, $N_{\Bs\to\Dstarzb\phi}$, and  $f_{\rm L}$. The coverage tests perform as expected, except for $N_{\Bs\to\Dzb\phi}$, for which the data uncertainty is overestimated by about 10\%. No correction is applied for this over-coverage.  While the fit is unbiased when using a single set of parameters to generate the $\Kp\Km$ background, when allowing for different true values of $a$ and $b$ in the different mass regions a bias on the parameter $N_{\Bz\to\Dzb\phi}$ is found and corresponds to an overestimation by 7 candidates.  This is corrected for the computation of the branching fraction.

Potential sources of systematic uncertainty on the efficiencies are correlated and largely cancel in the quoted ratios of branching fractions.  The main differences are related to the PID selection for the $\pip \pim$ and $\Kp \Km$ pairs and to the hardware trigger. For each effect, a systematic uncertainty of $2\%$ is computed, mainly from the PID calibration method and differences between the trigger response in data and simulation~\cite{LHCb-PAPER-2018-014}. The uncertainty on the known value of $\BR(\phi\to \Kp \Km)$ is $1\%$~\cite{PDG2018}. For the $\Bs$ modes, an uncertainty of $5.8\%$ related to the fragmentation factor ratio $f_s/f_d$~\cite{fsfd} is accounted for. The yield of the normalisation mode is assigned a systematic uncertainty of $2\%$, where the main contributions are from the modelling of the signal and partially reconstructed background shapes~\cite{LHCb-PAPER-2018-014}.

Sources of systematic uncertainty on the determination of $N_{\B^0_{(s)} \to \Dstarzbpar \phi}$ and $f_{\rm L}$ are related to the fit model of the $m_{\Kp\Km}$ distribution and that of the fit to the weighted $\Dzb\Kp\Km$ invariant-mass spectrum. The weights from the fits are calculated and the $\B^0_{(s)} \to \Dstarzbpar\phi$ yields and $f_{\rm L}$  are fitted with three different configurations: by varying the natural width of the $\phi$ meson by its uncertainty~\cite{PDG2018}; by replacing the quadratic part of the $m_{\Kp\Km}$ background PDF by a third-order Chebyshev polynomial; and by replacing the $m_{\Kp\Km}$ background PDF with an empirical function~\cite{PhysRevLett.98.211802}, ${(1-\exp(-\frac{m-m_0}{f})) \times (\frac{m}{m_0})^{c} + d \times (\frac{m}{m_0}-1)}$, where $m_0$ is fixed to $2m_K$ and the parameters $c$, $d$, and $f$ are free to vary in the fit. The largest variations from the nominal model are taken as systematic uncertainties. For the fit to the invariant-mass distribution of the $\Dzb\phi$ candidates, alternative models for $B^0_{(s)} \to\Dzb\Kp\Km$ and $\Bs \to \Dstarzb \phi$ are considered: one changing the fit model of the $B^0_{(s)} \to\Dzb\phi$ decays to that used to model $B^0_{(s)} \to\Dzb \Kp \Km$, as described in Ref.~\cite{LHCb-PAPER-2018-014}, and others in which the PDFs of the fully transversally/longitudinally polarised $\Bs\to\Dstarzb \phi$ decays are varied within the uncertainties on the ratio of branching fractions $\BR(\Dstarzb\to\Dzb \piz)/\BR(\Dstarzb\to\Dzb \gamma)$~\cite{PDG2018} and of the efficiencies obtained from simulation. 
Possible partially reconstructed background from the $\Bz \to \Dzb \phi \pi^+$ and $\Bs \to \Dzb \phi \pi^+$ decays are also considered in the fit model. 
The resulting uncertainties are summed linearly assuming maximal correlation for this kind of systematic uncertainty and correspond to relative values of $4.7\%$, $31.1\%$, $5.4\%$, and $4.9\%$ on $N_{\Bs \to \Dzb \phi}$, $N_{\Bz \to \Dzb \phi}$, $N_{\Bs \to \Dstarzb \phi}$, and $f_L$, respectively.   As the efficiencies depend on the signal decay-time distribution, the effect due to the different lifetimes of the $B_s^0$ eigenstates~\cite{PhysRevD.86.014027} is considered and found to be 0.8\%. When considering the ratio between $\BR(\Bs\to \Dstarzb \phi)$ and $\BR(\Bs\to \Dzb \phi)$ and the longitudinal polarisation fraction $f_L$, this systematic uncertainty is doubled to account for unknown strong phases between decay amplitudes and unknown fractions between different angular momentum. The systematic uncertainties from the various sources are listed in Table~\ref{tab:Systematics_DPhi_Summary}.
 \begin{table}[h]
\centering
\caption{Relative systematic uncertainties given in percent on the ratios of branching fractions and on longitudinal polarisation.}
\begin{tabular}{lccccc}
        \hline \hline
        \vspace{-0.3cm} \\
        Source & $\frac{\BR(\Bs\to \Dzb \phi)}{\BR(\Bz\to\Dzb \pip \pim)}$ & $\frac{\BR(\Bz\to \Dzb \phi)}{\BR(\Bz\to\Dzb \pip \pim)}$ & $\frac{\BR(\Bs\to \Dstarzb \phi)}{\BR(\Bz\to\Dzb \pip \pim)}$ & $\frac{\BR(\Bs\to \Dstarzb \phi)}{\BR(\Bs\to \Dzb \phi)}$ &$f_L$ \\
        \vspace{-0.3cm} \\
        \hline
        $N_{\B^0_{(s)} \to \Dstarzbpar \phi}$ & 4.7 & \alm31.1 & 5.4 & 6.4 &4.9 \\
        $N_{\Bz\to\Dzb \pip \pim}$ & 2.0 & 2.0 & 2.0 & $-$ & $-$ \\
        $\epsilon_{\textrm{PID}}$ & 2.0 & 2.0 & 2.0 & $-$  & $-$\\
        $\epsilon_{\textrm{trigger}}$ & 2.0 & 2.0 & 2.0 & $-$ & $-$\\
        $\BR(\phi\to \Kp \Km)$ & 1.0 & 1.0 & 1.0 & $-$  & $-$\\
        $f_s/f_d$ & 5.8 & $-$ & 5.8 & $-$ & $-$\\
        Lifetime &  0.8 & $-$ & 0.8 & 1.6 & $1.6$\\
        \hline
        Total & 8.3 & \alm31.2 & 8.8 & 6.6 & 5.2 \\
        \hline
\end{tabular}
\label{tab:Systematics_DPhi_Summary}
\end{table}

The ratio of branching fractions ${\BR(\Bs \to \Dzb \phi)}/{\BR(\Bz\to\Dzb \pip \pim)}$ is measured to be $(3.4 \pm 0.4 \pm 0.3)\%$, where the first uncertainty is statistical and the second systematic, and ${\BR(\Bs \to \Dzb \phi)}$ to be $(3.0 \pm 0.3 \pm 0.2 \pm 0.2) \times 10^{-5}$, where the third uncertainty is related to the branching fraction of the normalisation  mode~\cite{Kuzmin:2006mw,LHCb-PAPER-2014-070,PDG2018}. The branching fraction is compatible with and more precise than the previous \lhcb measurement~\cite{LHCb-PAPER-2013-035} and supersedes it. The decay  $\Bs\to\Dstarzb\phi$ is observed for the first time, with a significance of more than seven standard deviations estimated using its statistical uncertainty and systematic variations of $N_{\Bs\to\Dstarzb\phi}$.  The ratio of branching fractions ${\BR(\Bs\to\Dstarzb\phi)}/{\BR(\Bz\to\Dzb \pip \pim)}$ is measured to be  $(4.2 \pm 0.5 \pm 0.4)\%$ and the branching fraction $\BR(\Bs\to\Dstarzb\phi)$ is $(3.7 \pm 0.5 \pm 0.3 \pm 0.2) \times 10^{-5}$. The fraction of longitudinal polarisation is measured to be ${f_{\rm L} =(73 \pm 15 \pm 4)\%}$,  which is comparable with measurements from similar colour-suppressed $\Bd$ decays~\cite{Lees:2011gw,Matvienko:2015gqa}. The ratio of branching fractions ${\BR(\Bs\to\Dstarzb\phi)}/{\BR(\Bs\to\Dzb \phi)}$ is $1.23 \pm 0.20 \pm 0.08$.

The ratio of branching fractions of \mbox{${\BR(\Bz\to\Dzb\phi)}/{\BR(\Bz\to\Dzb \pip \pim)}$} is measured to be $(1.2 \pm 0.7 \pm 0.4) \times 10^{-3}$ and the branching fraction \mbox{$\BR(\Bz\to\Dzb\phi)$} to be \mbox{$(1.1 \pm 0.6 \pm 0.3 \pm 0.1) \times 10^{-6}$}. The significance for the $W$-exchange OZI-suppressed decay $\Bd \to \Dzb \phi$ is about two standard deviations. Since there is no significant signal, an upper limit is set as $\BR(\Bz\to\Dzb\phi) < 2.0 \ (2.3) \times 10^{-6}$ at $90\%$ (95\%) confidence level (CL), representing a factor of six improvement over the previous limit by the  BaBar collaboration~\cite{Aubert:2007nw}. The upper limit obtained here is compatible with the updated theoretical prediction ${\BR(\Bz\to\Dzb\phi) = (1.6 \pm 0.1)\times 10^{-6}}$. These results are used to constrain the $\omega - \phi$ mixing angle assuming the dominant contribution to the $\Bd \to \Dzb \phi$ decay is through $\omega - \phi$ mixing. The study in Ref.~\cite{Benayoun:2007cu} predicts a mixing angle between $0.45^\circ$ (at the $\omega$ mass) and $4.65^\circ$ (at the $\phi$ mass).
Using the upper limit in this Letter, the constraint $|\delta| < 5.2^\circ~ (5.5^\circ)$ is set at $90\%$ (95\%) CL. Further studies with more data are therefore motivated.

\section*{Acknowledgements}
%
%
\noindent We express our gratitude to our colleagues in the CERN
accelerator departments for the excellent performance of the LHC. We
thank the technical and administrative staff at the LHCb
institutes.
We acknowledge support from CERN and from the national agencies:
CAPES, CNPq, FAPERJ and FINEP (Brazil); 
MOST and NSFC (China); 
CNRS/IN2P3 (France); 
BMBF, DFG and MPG (Germany); 
INFN (Italy); 
NWO (Netherlands); 
MNiSW and NCN (Poland); 
MEN/IFA (Romania); 
MinES and FASO (Russia); 
MinECo (Spain); 
SNSF and SER (Switzerland); 
NASU (Ukraine); 
STFC (United Kingdom); 
NSF (USA).
We acknowledge the computing resources that are provided by CERN, IN2P3
(France), KIT and DESY (Germany), INFN (Italy), SURF (Netherlands),
PIC (Spain), GridPP (United Kingdom), RRCKI and Yandex
LLC (Russia), CSCS (Switzerland), IFIN-HH (Romania), CBPF (Brazil),
PL-GRID (Poland) and OSC (USA).
We are indebted to the communities behind the multiple open-source
software packages on which we depend.
Individual groups or members have received support from
AvH Foundation (Germany);
EPLANET, Marie Sk\l{}odowska-Curie Actions and ERC (European Union);
ANR, Labex P2IO and OCEVU, and R\'{e}gion Auvergne-Rh\^{o}ne-Alpes (France);
Key Research Program of Frontier Sciences of CAS, CAS PIFI, and the Thousand Talents Program (China);
RFBR, RSF and Yandex LLC (Russia);
GVA, XuntaGal and GENCAT (Spain);
Herchel Smith Fund, the Royal Society, the English-Speaking Union and the Leverhulme Trust (United Kingdom);
Laboratory Directed Research and Development program of LANL (USA).

%

\addcontentsline{toc}{section}{References}
\setboolean{inbibliography}{true}
\ifx\mcitethebibliography\mciteundefinedmacro
\PackageError{LHCb.bst}{mciteplus.sty has not been loaded}
{This bibstyle requires the use of the mciteplus package.}\fi
\providecommand{\href}[2]{#2}

 \newpage


\centerline{\large\bf LHCb collaboration}
\begin{flushleft}
\small
R.~Aaij$^{27}$,
B.~Adeva$^{41}$,
M.~Adinolfi$^{48}$,
C.A.~Aidala$^{73}$,
Z.~Ajaltouni$^{5}$,
S.~Akar$^{59}$,
P.~Albicocco$^{18}$,
J.~Albrecht$^{10}$,
F.~Alessio$^{42}$,
M.~Alexander$^{53}$
% ...

{\footnotesize \it
% ...
$ ^{5}$Clermont Universit{\'e}, Universit{\'e} Blaise Pascal, CNRS/IN2P3, LPC, Clermont-Ferrand, France\\
% ...
$ ^{10}$Fakult{\"a}t Physik, Technische Universit{\"a}t Dortmund, Dortmund, Germany\\
% ...
$ ^{18}$INFN Laboratori Nazionali di Frascati, Frascati, Italy\\
% ...
$ ^{27}$Nikhef National Institute for Subatomic Physics, Amsterdam, Netherlands\\
% ...
$ ^{41}$Instituto Galego de F{\'\i}sica de Altas Enerx{\'\i}as (IGFAE), Universidade de Santiago de Compostela, Santiago de Compostela, Spain\\
$ ^{42}$European Organization for Nuclear Research (CERN), Geneva, Switzerland\\
% ...
$ ^{48}$H.H. Wills Physics Laboratory, University of Bristol, Bristol, United Kingdom\\
% ...
$ ^{53}$School of Physics and Astronomy, University of Glasgow, Glasgow, United Kingdom\\
% ...
$ ^{59}$University of Cincinnati, Cincinnati, OH, United States\\
% ...
$ ^{61}$Syracuse University, Syracuse, NY, United States\\
% ...
$ ^{73}$University of Michigan, Ann Arbor, United States, associated to $^{61}$\\
% ...
}
\end{flushleft}

\end{document}